\documentclass[prb,twocolumn,showpacs,superscriptaddress,preprintnumbers]{revtex4}
\usepackage{graphicx}
\usepackage{dcolumn}
\usepackage{bm}

\input{tcilatex}

\begin{document}

\title{Strong-coupling Effects in cuprate High-$T_{c}$ Superconductors by
magnetooptical studies}
\author{Y. S. Lee}
\affiliation{Department of Physics, University of California at San Diego, La Jolla,
California 92093-0319, USA}
\author{Z. Q. Li}
\affiliation{Department of Physics, University of California at San Diego, La Jolla,
California 92093-0319, USA}
\author{W. J. Padilla}
\affiliation{Department of Physics, University of California at San Diego, La Jolla,
California 92093-0319, USA}
\author{S. V. Dordevic}
\affiliation{Department of Physics, Brookhaven National Laboratory, Upton, New York
11973, USA}
\author{C. C. Homes}
\affiliation{Department of Physics, Brookhaven National Laboratory, Upton, New York
11973, USA}
\author{Kouji Segawa}
\affiliation{Central Research Institute of Electric Power Industry, Komae, Tokyo
201-8511, Japan}
\author{Yoichi Ando}
\affiliation{Central Research Institute of Electric Power Industry, Komae, Tokyo
201-8511, Japan}
\author{D. N. Basov}
\affiliation{Department of Physics, University of California at San Diego, La Jolla,
California 92093-0319, USA}
\date{\today }

\begin{abstract}
Signatures of strong coupling effects in cuprate high-$T_{c}$
superconductors have been authenticated through a variety of spectroscopic
probes. However, the microscopic nature of relevant excitations has not been
agreed upon. Here we report on magneto-optical studies of the CuO$_{2}$
plane carrier dynamics in a prototypical high-$T_{c}$ superconductor YBa$%
_{2} $Cu$_{3}$O$_{y}$ (YBCO). Infrared data are directly compared with
earlier inelastic neutron scattering results by Dai \textit{et al}. [Nature
(London) \textbf{406}, 965 (2000)] revealing a characteristic depression of
the magnetic resonance in H $\parallel $ \textit{c} field less than 7 T.
This analysis has allowed us to critically assess the role of magnetic
degrees of freedom in producing strong coupling effects for YBCO system.
\end{abstract}

\pacs{74,25.Gz, 74,72.Bk}
\maketitle

Electron pairing in conventional superconducting metals is mediated by
phonons. Strong interaction of electrons with the lattice also manifests
itself through self-energy effects yielding fingerprints of the
electron-phonon spectral function $\alpha ^{2}F(\omega )$ in the tunneling
density of states,\cite{carbotte-rmp} infrared (IR) conductivity\cite%
{richards-67,timusk76} or energy band dispersion probed in the angle
resolved photoemission spectroscopy (ARPES).\cite{damascelli03,alpha} A
similar clear understanding of the cuprate high-$T_{c}$ superconductors is
yet to be achieved. Spectroscopic probes of self-energy unequivocally prove
the relevance of strong coupling effects.\cite%
{damascelli03,Puchkov96,zasadzinski01} However the microscopic origin of the
pertinent spectral function is still debated. Numerous experiments are
suggestive of quasiparticle (QP) interaction with a magnetic resonance mode%
\cite{norman,carbotte99,Munzar99,johnson01,abanov99,coupling,Hwang04} seen
in inelastic neutron scattering (INS) experiments.\cite{dai01,fong00} An
issue of whether or not the magnetic mode is capable of having a serious
impact on the electronic self-energy, in view of the small intensity of the
resonance, has been contested on theoretical grounds.\cite%
{abanov02,kee02,chubukov04} Moreover, a recent re-examination of ARPES
results\cite{lanzara01,lanzara04,devereaux04} has suggested that the
totality of data is better described in terms of coupling to phonons and not
to magnetic excitations. However, this latter claim is not supported by IR
studies of isotopically substituted YBa$_{2}$Cu$_{3}$O$_{y}$ (YBCO) which
show no isotope effect for the feature in question.\cite{wang02,bernhard04}
Thus currently available data leave ambiguities regarding the roles of
lattice and magnetic degrees of freedom in carrier dynamics as well as in
the superconductivity of cuprates.

Insights into strong coupling effects may be gained from studies of the QP
dynamics in magnetic field. The rationale for this approach is provided by
the work of Dai \textit{et al}. who discovered that the intensity of the
magnetic resonance in the $y$ = 6.6 YBCO crystal ($T_{c}$ = 62.7 K) is
suppressed by 20 $\%$ in 6.8 T field applied along the $c$-axis.\cite{dai00}
Other candidate excitations including phonons, or the continuum of spin
fluctuations, are unlikely to be influenced by a magnetic field of similar
modest magnitude. For this reason an exploration of the field-induced
modifications of the electronic self-energy enables a direct experimental
inquiry into the role of the magnetic resonance in QP properties and on a
more general level, into an intricate interplay between superconductivity
and magnetism in cuprates. Here we report on studies of a magnetic field
dependent QP response in a series of YBCO crystals using IR spectroscopy.
Changes of the optical conductivity and of the $\alpha ^{2}F(\omega )$
spectrum extracted from the data in 7 T field are found to be within the
uncertainty of our measurements. This null result nevertheless allows us to
critically assess the strong coupling scenario in high-$T_{c}$
superconductors. 
\begin{figure}[tbp]
\includegraphics[width=0.38\textwidth]{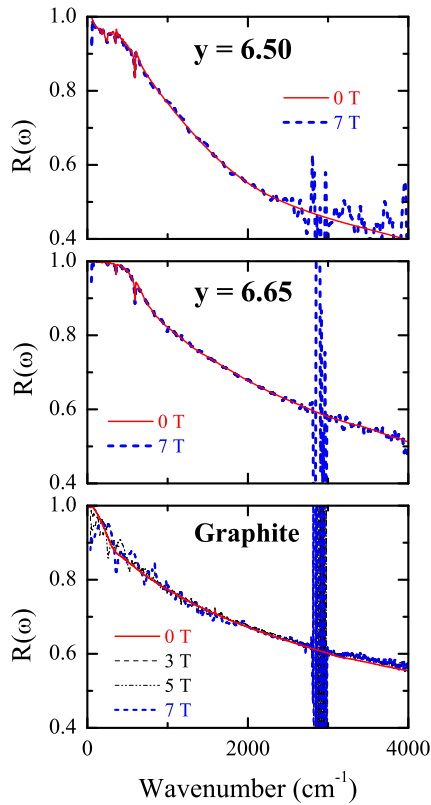}
\caption{(color online) Reflectance spectra obatined at 5 K in magnetic
field for (top) $y$ = 6.50 ($T_{c}$ $\sim $ 31 K), (middle) $y$ = 6.65 ($%
T_{c}$ $\sim $ 60 K) YBCO crystals, and (bottom) graphite. Polarized light
along the $a$ axis is used for detwinned YBCO crystals. The magnetic field
is applied along the $c$ axis. Red (thin solid) lines: \textbf{H} = 0; blue
(thick dashed) lines: \textbf{H }= 7 T. Sharp spikes near 2,900 cm$^{-1}$ in
the high field spectra are due to absorption in the windows of our cryostat. 
}
\end{figure}

We investigated YBCO single crystals with $y$ = 6.50 ($T_{c}$ $\sim $ 31 K)
and 6.65 ($T_{c}$ $\sim $ 60 K) grown by a conventional flux method and
detwinned under uniaxial pressure at CRIEPI.\cite{Segawa01} Detwinned
crystals are necessary to perform reflectivity measurements with the
polarization of IR light along the $a$ axis. In this geometry the properties
studied here unveil the intrinsic dynamics of the CuO$_{2}$ planes without
contamination by the chain segments extending along the $b$ axis.\cite%
{chains} Field-dependent reflectivity spectra were recorded at UCSD in the
frequency range 20 - 5000 cm$^{-1}$,\cite{magnetooptic} and were
supplemented by zero field reflectance and ellipsometry data up to 5 eV. The
magnetic field aligned along the $c$ axis was applied using a
superconducting split-coil magnet. Our magneto-optical apparatus enables
absolute measurements of reflectivity. The 0 T data obtained using this
instrument were found to be consistent with the spectra obtained by means of
our compact reflectometer for temperature-dependent reflectance. Owing to
the small cyclotron frequency of YBCO it is appropriate to extract the
complex conductivity $\widetilde{\sigma }(\omega )=\sigma _{1}(\omega
)+i\sigma _{2}(\omega )$ from reflectance spectra using the standard
Kramers-Kronig equations.

Representative results are displayed in Fig. 1. Here we plot the raw
reflectance spectra measured at $T$ = 5 K for $y=6.50$ and $6.65$ crystals.
The spectra for the latter material are in good agreement with the earlier
studies of YBCO with similar oxygen content.\cite{basov96} Notably, we found
that the field-induced changes of the reflectivity are negligibly small
either under zero-field cooling or under in-field cooling conditions. In
order to quantify the experimental accuracy achievable in our
magneto-optical apparatus, in the bottom panel of Fig. 1 we also show
spectra for highly oriented pyrolytic graphite (HOPG) measured with the
polarization of the \textbf{E} vector along the graphene sheets. A
comparison is warranted by the similarity in the zero-field optical
properties of graphite and high-$T_{c}$ cuprates. In the former system we
are capable of resolving small (less than 1 $\%$) changes of the overall
reflectivity level as well as weak structure associated with the Landau
level transitions triggered by the magnetic field perpendicular to grapheme
planes.\cite{graphite} No such changes are detectable for YBCO.

We proceed by briefly outlining the fundamentals of an IR probe of the
electronic self-energy. Interaction of the mobile charges with bosonic
excitations leads to a frequency dependence of the scattering rate $1/\tau
(\omega )$ in accord with the Allen formula:\cite{allen71} 
\begin{equation}
\frac{1}{\tau (\omega )}=\frac{2\pi }{\omega }\int_{0}^{\omega }d\omega
^{\prime }(\omega -\omega ^{\prime })\alpha ^{2}F(\omega ^{\prime })+\frac{1%
}{\tau _{\text{imp}}}\text{,}  \label{eq:one}
\end{equation}%
where $1/\tau _{\text{imp}}$ is the impurity scattering. Experimentally, the
frequency dependence of $1/\tau (\omega )$ can be inferred from the analysis
of the complex optical conductivity $\widetilde{\sigma }(\omega )$ within
the Extended Drude Model:\cite{Puchkov96} $1/\tau (\omega )=\omega
_{p}^{2}/4\pi \cdot \func{Re}[1/\sigma (\omega )]$, where a total plasma
frequency $\omega _{p}^{2}$ is determined by integration of $\sigma
_{1}(\omega )$ up to the charge transfer gap. Eq.(1) is commonly applied to
the analysis of the data for cuprates and provides support for an idea of
QPs coupling to a magnetic resonance.\cite{carbotte99,Hwang04} Nevertheless,
Eq.~(1) is not entirely adequate for \textit{a superconductor} since it
completely ignores the effect of the superconducting energy gap $2\Delta $
on the form of the $1/\tau (\omega )$ spectra. In order to treat the impact
of the gap and of strong coupling to bosonic modes on equal footing we used
the following result also derived by Allen:\cite{allen71} 
\begin{eqnarray}
\frac{1}{\tau _{s}(\omega )} &=&\frac{2\pi }{\omega }\int_{0}^{\omega -2\Delta
}d\omega ^{\prime }(\omega -\omega ^{\prime })\alpha ^{2}F(\omega ^{\prime
}) \nonumber \\ 
&&\times E\left[ \left( 1-\frac{4\Delta ^{2}}{(\omega -\omega ^{\prime })^{2}%
}\right) ^{1/2}\right] \text{,}
\end{eqnarray}%
where $E$ is the complete elliptic integral of second kind. Although the
utility of Eq. (2) is obvious it is non-trivial to employ this formula for
the extraction of $\alpha ^{2}F(\omega )$ from experimental data since
simple inversion prescriptions do not apply in this case. To circumvent this
limitation Dordevic \textit{et al.} developed a numerical procedure based on
the inverse theory that is described in details elsewhere.\cite{dordevic04}
In the bottom panels [(e) and (f)] of Fig.~2 we show the $\alpha
^{2}F(\omega )$ spectrum extracted in this fashion from the \textbf{H} = 0
spectrum. We wish to point out an excellent agreement with INS results for
the spin susceptibility $\chi (\omega )$ [open symbols in Fig. 2(e)]\cite%
{dai99} without introducing a frequency offset.\cite{carbotte99}\ Indeed,
both a sharp resonance and a broad incoherent background of the spin
susceptibility appear to be reproduced in the $\alpha ^{2}F(\omega )$\
spectrum.\cite{negative}

\begin{figure}[tbp]
\includegraphics[width=0.45\textwidth]{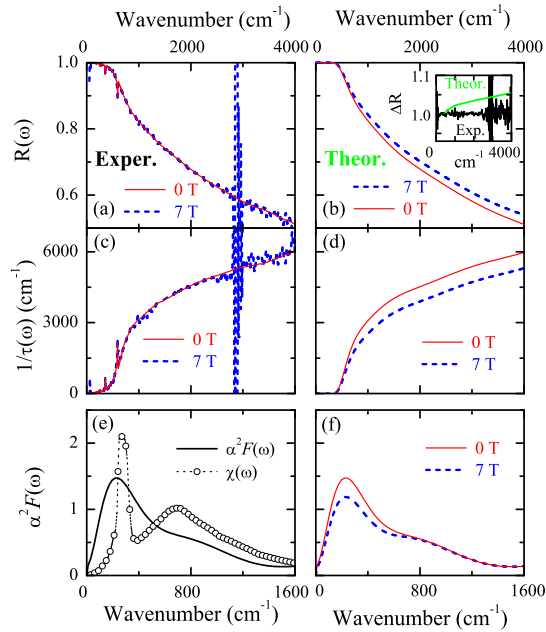}
\caption{(color online) Low temperature reflectance spectra $R$($\protect%
\omega $), 1/$\protect\tau (\protect\omega )$ spectra and $\protect\alpha %
^{2}F(\protect\omega )$ data for $y$ = 6.65 YBCO single crystal. Red (thin
solid) lines: \textbf{H} = 0; blue (thick dashed) lines: \textbf{H }= 7 T.
Left panels: experimental results. Right panels: model spectra calculated
using the protocol described in the text. Inset in (b): $\Delta R(\protect%
\omega ,$\textbf{H}$)=R(\protect\omega ,7$ T)/$R$($\protect\omega $, 0 T).
Sharp spikes in the high field spectra are due to absorption in the windows
of our cryostat. To calculate $\protect\alpha ^{2}F(\protect\omega )$ we
used $\Delta =180$ cm$^{-1}$ in Eq. (2). Also shown with open symbols in
panel (e) is the spin susceptibility $\protect\chi (\protect\omega )$ from
the INS data reported in Ref. \protect\cite{dai99} for $y$ = 6.6 ($T_{c}$ =
62.7 K) single crystal. The $\protect\chi (\protect\omega )$ spectrum is
similar to the experimental result for $\protect\alpha ^{2}F(\protect\omega %
) $ obtained from the inversion of IR data.}
\end{figure}

An important feature of the strong coupling formalism [Eqs.(1) and (2)] is
the integral relationship between $1/\tau (\omega )$ and $\alpha
^{2}F(\omega )$. This relationship implies that a depression of the
intensity in $\alpha ^{2}F(\omega )$ necessarily reduces the magnitude of $%
1/\tau (\omega )$ and consequently enhances the reflectivity level at all
frequencies \textit{above} the resonance mode in the spectral function. In
order to quantify the magnitude of possible \textbf{H}-induced changes
associated with a depression of the INS resonance in magnetic field we
adopted the following protocol. We first reduced the intensity of the sharp
peak near $\sim $ 270 cm$^{-1}$ ($\sim $ 34 meV) \ in the $\alpha
^{2}F(\omega )$ spectrum by 20 $\%$: a factor suggested by INS measurements.%
\cite{dai00} The intensity of broad background remained intact [blue (thick
dashed) line in Fig. 2(f)]. Evidently, this modification will produce a
conservative estimate of the impact of the INS resonance on IR data. Using
the spectral function with the suppressed intensity we calculated $1/\tau
(\omega ,7$ T$)$ from Eq. (2) and also $m^{\ast }(\omega $, 7 T$)$ with the
help of Kramers-Kronig analysis. Finally, a combination of $1/\tau (\omega
,7 $ T$)$ and $m^{\ast }(\omega $, 7 T$)$ allowed us to generate the
reflectance spectrum $R(\omega ,$ 7 T$)$ [blue (thick dashed) line in Fig.
2(b)]. Comparing this final output of modeling with the experimental curve
for \textbf{H} = 0 one finds that the effect of the applied magnetic field
is rather small in the far-IR but exceeds 5 $\%$ at frequencies above 800 cm$%
^{-1}$. This is further detailed in the inset of Fig. 2 where we present the
ratio $\Delta R(\omega ,$\textbf{H}$)=R(\omega ,7$ T)/$R$($\omega $, 0 T)
calculated from the model spectra. These anticipated changes of reflectance
exceed the uncertainty of $R(\omega ,$\textbf{H}$)$ in our apparatus and
therefore should be readily detectable.

Empowered by modeling of the data we will now discuss the implications of
the lack of magnetic field dependence of IR spectra for underdoped YBCO\
documented in Figs.~1 and 2. One possible interpretation of the data is that
the magnetic resonance is irrelevant to QP dynamics. Within this view
self-energy effects in the data can be assigned to excitations inherently
insensitive to the magnetic field such as phonons or the spin fluctuations
continuum.\cite{chubukov} However, single-phonon processes have a well
defined high-energy cut-off in cuprates that does not exceed 800 cm$^{-1}$
for YBCO. For this reason phonons alone cannot account for a high frequency
background in the $\alpha ^{2}F(\omega )$ spectra in Fig. 2. On the
contrary, magnetic excitations extend to significantly higher frequencies
and therefore can naturally account for the form of $1/\tau \left( \omega
\right) $ spectra in mid-IR energy range. Thus our results are consistent
with the viewpoint that distinct phonon modes in concert with the broad spin
fluctuations continuum are jointly responsible for strong coupling effects
in cuprates.

An intriguing interpretation of the magnetic resonance seen in the INS
experiments is offered by $SO$(5) theory also providing a unified view on
superconductivity and antiferromagnetism in cuprates.\cite%
{demler95,zhang97,demler04} This interpretation is in accord with our data
as we will elaborate below. $SO$(5) theory predicts a $\pi $-resonance in
the particle-particle channel that is present both above and below $T_{c}$.
Coupling of the $\pi $-resonance to neutrons is facilitated by the formation
of the pair condensate in a $d$-wave superconductor. This latter attribute
of the $\pi $-mode is important. First, it allows one to understand the
quasiparticles self-energy effects at $T>T_{c}$ in the absence of
superconductivity. Second, within the framework of the $SO$(5) theory a
suppression of the neutron mode in the high magnetic field INS experiments
is only an apparent effect. Indeed, this suppression is fully accounted for
by a reduction of the superconducting order parameter in type-II $d$-wave
system that is expected to occur in the regime of constant intrinsic
intensity of the $\pi $-mode. In this fashion magnetic field is expected to
have only a small effect on the quasiparticles self-energy probed in the IR
data despite apparent depression of the mode seen in the INS experiment.
Thus the $SO$(5) interpretation of the INS peak allows one to reconcile
dissimilarities in the magnetic field effects in IR and neutron measurements.

The results reported here call for an examination of the self-energy effects
seen in cuprates by other spectroscopic methods in magnetic field. While it
may be impossible to carry out such experiments in the case of photoemission
studies, tunneling measurements appear to be well suited for this task. It
is also worthwhile to re-evaluate the role of interband transitions and
other excitations in providing a direct contribution to the optical
conductivity in mid-IR region. The so-called multi-component analysis of the
optical data offers a complementary interpretation of some of the effects
discussed here within the self-energy formalism.\cite{quijada99,yslee05}

We acknowledge G. Blumberg, K.S. Burch, J.P. Carbotte, A. Chubukov, E.
Demler, S. Kivelson, T. Timusk, and J.M. Tranquada, S.C. Zhang for helpful
discussions. This research was supported by the US department of Energy
Grant and NSF.

\newpage


\begin{thebibliography}{99}
\bibitem{alpha} The Eliashberg spectral function $\alpha^2F(\omega)$
accounts for an effective (and generally energy dependent) electron-phonon
coupling $\alpha^2$; $F(\omega)$ is the phonon density of states.
\bibitem{negative} One distinction of the spectral function obtained using
the inversion theory from that generated using a simplified procedure $%
W(\omega )=1/2\pi \cdot d^{2}/d\omega ^{2}[\omega /\tau (\omega )]$ is
worthy of attention. The latter approach always yields regions of negative
spectral function which is unphysical. On the contrary, the inversion theory
offers a self-consistent treatment of both energy gap and strong coupling
effects in the optical data and yields the physically meaningful spectral
function throughout the entire frequency range. For details, see Ref. \cite%
{dordevic04}.
\end{thebibliography}
\end{document}